\documentclass[referee]{aastex}
\usepackage{spr-astr-addons}

\usepackage{graphicx}
\usepackage{lscape}

\newcommand{\argonone}{Ar~{\footnotesize{I}}}   
\newcommand{\carthree}{C~{\footnotesize{III}}}  
\newcommand{\hone}{H~{\footnotesize{I}}}        
\newcommand{\nittwo}{N~{\footnotesize{II}}}      
\newcommand{\oxyeight}{O~{\footnotesize{VIII}}} 
\newcommand{\oxyseven}{O~{\footnotesize{VII}}}  
\newcommand{\oxysix}{O~{\footnotesize{VI}}}     
\newcommand{\ironnine}{Fe~{\footnotesize{IX}}}        
\newcommand{\ironten}{Fe~{\footnotesize{X}}}        
\newcommand{\ironeleven}{Fe~{\footnotesize{XI}}}        
\newcommand{\sodiumone}{Na~{\footnotesize{I}}}        
\newcommand{\silthree}{Si~{\footnotesize{III}}} 

\newcommand{\chandra}{{\it Chandra}}

\newcommand{\copernicus}{{\it Copernicus}}

\newcommand{\rosat}{{\it ROSAT}}
\newcommand{\suzaku}{{\it Suzaku}}
\newcommand{\xmm}{{\it XMM-Newton}}

\shorttitle{Trouble with theBubble}
\shortauthors{Welsh $\&$ Shelton}

\begin{document}


\title{The Trouble with the Local Bubble}


\author{Barry Y. Welsh\altaffilmark{1}
and
Robin L. Shelton\altaffilmark{2}}

\altaffiltext{1}{Space Sciences Laboratory, University of California, 7 Gauss Way, Berkeley, CA 94720, USA}
\altaffiltext{2}{Department of Physics and Astronomy, University of Georgia, Athens, GA 30602, USA}

\begin{abstract}
The
model of a Local Hot Bubble has been widely accepted as providing a framework
that can explain the ubiquitous presence of the soft X-ray background diffuse emission.
We summarize the current knowledge on this local interstellar region, paying particular reference to
observations that sample emission from the presumed local million degree K hot plasma.
However, we have listed numerous observations that are seemingly in conflict with the
concept of a hot Local Bubble. In particular, the discovery of solar wind charge exchange
that can generate an appreciable soft X-ray background signal within the heliosphere, has led to a re-assessment
of the generally accepted model that requires a hot local plasma.

In order to explain the majority of observations of the local plasma, we forward two new speculative
models that describe the physical state of the local interstellar gas.
One possible scenario is similar to the present widely accepted model of the Local Hot Bubble, except that it 
accounts for only 50$\%$ of the soft X-ray emission currently detected in the 
galactic plane, has a lower thermal pressure than previously thought, and its 
hot plasma is not as hot as previously believed. 
Although such a model can 
solve several difficulties
with the traditional hot Local Bubble model, a heating mechanism for the dimmer and 
cooler gas remains to be found.
The second possible explanation is that of the
`Hot Top'
model, in which the Local Cavity is an 
old supernova remnant  in which no (or very little) million degree local plasma is presently required.
Instead, the cavity is now thought to be filled with partially ionized cloudlets of temperature
$\sim$ 7000~K that are surrounded by lower density envelopes of photo-ionized gas
of temperature $\sim$ 20,000~K. Although this new scenario provides a natural
explanation for many of the observations that were in conflict with the Local Hot Bubble model, we
cannot (as yet) provide a satisfactory explanation or the emission levels observed in
the B and Be ultra-soft X-ray bands.
\end{abstract}


\section{Introduction}

The stellar winds and supernovae events that are associated with clusters of massive early-type 
stars have a  profound effect on the surrounding interstellar medium (ISM), including the creation 
of large cavities.  These cavities, which are often referred to as ``interstellar bubbles'', 
are typically $\sim$100~pc in diameter and have low neutral gas densities of
$n(H) \sim 0.01$~cm$^{-3}$ \citep{weave77}. They are
ubiquitously present both within our own 
Milky Way galaxy (e.g.  the Eridanus, Cygnus, and Loop I superbubbles) and within the Large and 
Small Magellanic Clouds (e.g. N44 and N19).  It is therefore unsurprising to learn that the Sun 
finds itself located within such a rarefied interstellar cavity \citep{welsh98, sfeir99, lall03}, which we shall subsequently refer to as 
the Local Cavity. 
The Sun's placement within a region of unusually low gas density has been 
known for over 40 years and was initially deduced from the very low values of interstellar 
extinction measured for stars within a distance of $\sim100$~pc, as compared with the reddening 
derived for the more distant regions of the galaxy \citep{fitz68,lucke78}.  
In addition, the small \hone\ column densities derived from Ly-alpha absorption measurements of nearby 
stars \citep{bohl75}and the lack of spectral hardening in the observed soft X-ray background signal, 
which requires a low value of local neutral gas absorption \citep{bowyer68}, also added 
credence to the belief that the solar neighborhood resides within a region of low neutral hydrogen 
density.

During the 1970's observations of a ubiquitous 0.25 keV soft X-ray background
signal by the 
Wisconsin group \citep{sanders77, will74} and the detection of the \oxysix\ (1032) 
absorption line towards several OB stars located within the solar neighborhood by the \copernicus\ 
satellite \citep{jenk78} provided strong supporting evidence that the Local Cavity was filled with a highly 
ionized and hot (million degree K), low density plasma. In addition, new theoretical models of the 
general ISM predicted the presence of a pervasive hot gas-phase component \citep{cox74, mckee77}.
Given that other superbubbles, such as those of Cygnus \citep{cash80} and the
Orion-Eridanus region \citep{reyn79}, were also being detected through their X-ray emission properties and that 
Heiles \citep{heiles79} had just compiled an extensive list of galactic bubbles/supershells observed in the 
radio 21cm \hone\-line, the observational and theoretical basis for believing that hot gas resided in 
the cavity seemed well substantiated \citep{cox87}.
This hot interstellar plasma was dubbed the 
Local Hot Bubble.

At the time of the the Local Bubble $\&$ Beyond Conference \citep{breit98}
the widely accepted picture was one in which the Sun is located within an extremely low 
neutral density ($n(HI) \sim 0.005$~cm$^{-3}$), high thermal temperature ($T \sim 10^{6}$~K) and 
highly ionized local plasma with dimensions of at least 80pc in many galactic directions. This high 
temperature plasma produces half to all of the observed soft X-ray background intensity in most galactic directions.  
The hot gas seen in low latitude directions must be local because there are nearby optically thick 
walls of \hone\ (the neutral boundary to the Local Cavity) towards which a non-zero 1/4 keV flux is observed 
\citep{snow00}. Lying within the rarefied and hot bubble 
are many low density 
($N(HI) \sim 0.1$~cm$^{-3}$), partially ionized ($n_e \sim 0.1$cm$^{-3}$) and warm  
($T \sim 7000$~K) cloudlets, the most well-studied being the local interstellar cloud,
within which the Sun is thought to reside \citep{lall92}. The origin of the Local Hot Bubble was 
thought to be linked to an explosive event in the nearby ($d \sim 150$~pc)  Sco-Cen OB association 
that occurred some 2~Myr ago \citep{frisch81}, or one or more nearby SN \citep{cox82,edgar93},
such that the local region can essentially be considered as an old cooling SNR or 
superbubble structure.

The observed anticorrelation of the soft X-ray background emission intensity with the total sight-line neutral hydrogen column density led to two main models that attempted to explain the observed spatial distribution of the X-ray emitting gas. In the absorption model the ubiquitous soft X-ray background emission was thought to originate outside the Galaxy and to be partially absorbed by Galactic material, although the absorption cross sections could not produce
 the observed dynamic ranges \citep{mccam76} or band ratios \citep{bloch86,juda91}. In the widely accepted displacement model of the Local Hot Bubble \citep{sanders77, cox87}, the observed soft X-ray background brightness was thought to be proportional to the path length through the hot Local Bubble gas, which is roughly anticorrelated with the total hydrogen gas column density. Although neither model could fully explain all of the observations of the soft X-ray background, the displacement model was stronger.  The prevailing thought was that the soft X-ray background emission signal observed along any sight-line is composed of 4 components: (i) emission from the Local Hot Bubble of interstellar plasma located in the galactic disk near to the Sun, (ii) emission from a more distant (halo) source of soft X-rays located at high galactic latitudes, (iii) emission from an isotropic extragalactic background and (iv) an emission component arising in the Galactic disk at distances beyond the absorbing wall of the Local Cavity \citep{kuntz00}.

Such a model could account for the relative scarcity of neutral gas  and dust for interstellar sight-lines that extend $< 100$~pc \citep{frisch83}. 
Furthermore, cold and dense high latitude clouds could block emission originating from a more distant halo
and/or extragalactic origin, thus creating observable `shadows' in the maps of the diffuse
soft X-ray background emission \citep{burrows91,snow93}.
By analyzing these shadows, it has been possible to isolate the contribution of foreground Local
Bubble emission from more distant emission.  
These shadowing experiments have provided the
strongest evidence for a million degree plasma located within a hot Local Bubble, with a 
typical contribution of $\sim 400 \times 10^{-6}$ R12  (i.e. \rosat\ R1 and R2 band) counts s$^{-1}$ arcmin$^{-2}$ to the total 1/4~keV soft X-ray background signal 
in most directions
\citep{snow00}.

We discuss additional properties of the hot local gas in Section 2.  
Several discrepancies
between the model of the Local Hot Bubble and the observations are also discussed
in this section.  The most difficult problem is that posed by the discovery
of solar wind X-rays, which appear to account for a significant
fraction of the flux that had been attributed to a million degree local plasma.  Taking into
account the solar wind X-ray  contribution, we
outline possible revisions to the Local Hot Bubble model in Sections 3
$\&$ 4.
Section 3 describes the least drastic revision, 
that in which the emission from a hot local plasma is
simply dimmer in soft X-rays than previously believed.    
Section 4
describes the other extreme, a new model called the Hot Top Model,
in which the solar wind X-rays account for all of the local X-rays in
the Galactic plane.

\section{Details of, and Problems with, the Local Hot Bubble Model}
In the following subsections we briefly discuss each of the major physical properties of the standard 
Local Hot Bubble model (i.e. temperature, pressure, size, etc).
Over the past decade observational evidence has emerged that has raised serious questions concerning the standard
 model of a hot local plasma.
Therefore, we also make note of observations that would seem to contradict several of the shibboleths concerning how the local ISM is currently perceived.

\subsection{ The Spectrum of the $10^6$~K Plasma}

Diffuse emission arising from a seemingly ubiquitous 0.25 keV soft X-ray background 
(soft X-ray background) signal has 
been widely observed in the Be, B, and C
 bands during numerous suborbital sounding rocket flights by the 
\citep{mccam90,mccam02} and during the \rosat\ soft X-ray background all-sky
survey \citep{snow98}. The galactic distribution of this emission, as recorded by
this latter survey,
is shown in Figure 1.

In the Galactic plane, apart from directions towards SNRs, superbubbles, or 
the Galactic center, observations
record only the emission that originates 
within the Local Hot Bubble. X-ray shadowing observations
using high latitude clouds
have also isolated this hot and local component.
The observed ratio of
flux in the B band relative to that in the C band and the ratio of
flux in the \rosat\ R1 band relative to
that in the \rosat\ R2 band are indicators of the plasma's temperature and
are roughly constant across the sky.
The prevailing explanation for these observations is that the 
X-ray emission arises in a pool of hot gas ($T \sim 10^6$~K)
which surrounds the Sun and whose radius varies with direction but whose properties do not.
There is a 2-sigma upper limit on the intermixed \hone\ column density within the Local Hot Bubble of $N(HI) < 6.6  \times 10^{18}$ cm$^{-2}$ \citep{juda91}.
This limit is in agreement with the inferred average \hone\ column density throughout the Local Cavity as derived from \sodiumone\ observations along sight-lines within 60pc \citep{welsh94}.
However, as discussed in Section 2.4, the shape of the
Local Hot Bubble, which is calculated from the assumption of uniform emissivity, is difficult to
reconcile with the shape of the surrounding structure.
The \rosat\ and Wisconsin X-ray broad-band data are generally
consistent with collisional ionizational equilibrium (CIE) emission models for a gas at
1 million $^{\
m{o}}$K.
Using higher resolution spectra of the hot emitting gas
the  \chandra , \xmm, and \suzaku\ satellites have
also found that the
data in the 0.4 to 0.7~keV band can be fit with CIE model spectra
\citep{mark03, gal07, hen07}.The actual
spectrum of the soft X-ray background in this energy range is seen
 to be dominated by emission from the \oxyseven\ and \oxyeight\ lines.
However,
CIE models that generate the observed surface brightness of the \oxyseven\ line are found to predict
too much lower energy 1/4 keV band emission for the MBM 12 direction
\citep{smith07}.

In a few cases theoretical models in which the X-ray emitting gas was slightly out of equilibrium such that the plasma is
slightly underionized could not be disallowed \citep{smith05, hen07}.
We note that models having a
highly non-equilibrium ionization (NEI) condition in which the plasma
is thought to be strongly over-ionized due to a previous event such as a supernova \citep{breit99, breit01}
are seemingly disallowed by several observations.
Recent $\it FUSE$ observations of the emission intensity of both \oxysix\ \citep{shelt03} and \carthree\ \citep{welsh03} cannot be reconciled with the NEI model of the Local Hot Bubble \citep{breit01} which implies line intensities far greater than those which are observed.

Additionally, observations of the diffuse extreme ultraviolet (EUV) sky background spectral emission using both the $\it CHIPS$ 
\citep{hur05}  and $\it EURD$ \citep{edel01} satellites are also in contradiction with
the predictions of the NEI model.
We also note that strong Si III (1206\AA) interstellar absorption has
been reported for several
sight-lines towards nearby stars \citep{hol99,nehme08}. This ion has a relatively high
rate coefficient for charge exchange with neutral hydrogen and 
its presence within the local interstellar medium is unlikely to be linked to the remnant of some past ionization event 
such as a supernova shock \citep{lyu96, breit99}.
 Depending on the details of the local ionizing radiation field, a
 high hydrogen ionization fraction (95$\%$) is required in order to sustain the presently
observed amounts of Si III. 

Two sets of high spectral resolution observations of the soft X-ray background have been made thus
far \citep{sand01,mccam02}.
The 148 to 284 eV observations by Sanders et al.
found a blend of emission lines, whose spectrum cannot
be reproduced by a solar abundance CIE plasma. 
This energy region is the most sensitive to the effects
of absorption and the observed region was in the galactic plane,
therefore the emission must originate in a local source of hot gas.
We note that a major difficulty with the CIE emission model is that
a million degree local plasma should also be bright at extreme ultraviolet (EUV) wavelengths,
but attempts to detect the associated EUV line emission have not been successful 
\citep{jel95,vall98,hur05}.  Although adherents to a Local Hot Bubble can explain these non-detections as being due to the local hot gas having a very low gas-phase iron abundance (thus making the detection of the expectedly strong \ironnine\ line at 171\AA\ problematic), we note that
 that Fe is unlikely to be depleted in typical hot ISM plasma \citep{yao06}..

\subsection{Cool Gas Inside and Outside the Local Hot Bubble}
The bubble of local hot gas is assumed
to fit within the Local Cavity, whose location is 
defined by the ``walls'' of cold and neutral material surrounding the local region.
The size and shape of the Local Cavity is best estimated from high spectral resolution optical studies of the NaI D-lines (5890) seen in absorption towards many sight-lines within 300pc. The NaI ion is a good tracer of cold ($T < 1000$~K) and neutral ($I.P. < 5.1$~eV) interstellar gas and has been widely used as a surrogate for local \hone\ absorption \citep{ferl85, welsh94}. In a recent NaI absorption survey of 1005 sight-lines within 250pc \citep{lall03}, 3-D maps of the density distribution of cold gas have revealed the Local Cavity to be an irregularly shaped rarefied region of  50 - 80pc radius in most directions in the Galactic plane.
Figure 2 shows that the Local Cavity is connected to adjacent low density interstellar regions (such as the Pleiades bubble and Loop III) through narrow interstellar tunnels, as predicted by theoretical models of the ISM \citep{cox74}.  These maps of the distribution of cold and neutral local gas reveal 3 interesting interstellar features: (i) a 200pc long by 50pc wide extension of the Local Cavity towards the direction of the star Beta CMa, (ii) extensions of the Local Cavity into the lower galactic halo through both of the open-ends of a rarefied Local Chimney feature \citep{welsh99}, see Figure 3 and (iii)  a connection through a disturbed interstellar region that opens into the Loop I superbubble at a distance of $\sim90$~pc \citep{welsh05}. We shall return later to the significance these latter two features in Section 3.

UV and optical absorption measurements towards stellar targets within 100pc of the Sun have
revealed the presence of many partially ionized and warm diffuse cloudlets located within
the Local Cavity \citep{lall03, red08}. These cloudlets
typically have hydrogen column densities, log N(HI) $\sim$ 10$^{17 - 18}$ cm$^{-2}$ and
temperatures of $<$ 7000~K and possess bulk motions generally differing by only 5 - 10 km s$^{-1}$.
Their presence within the Local Hot Bubble is thought to provide natural sites for transition temperature layers where ions such as \oxysix\ could reside \citep{slav02}. The most well-studied of these cloudlets is the local interstellar cloud,  partially within which
 the Sun is thought to reside.
The cloudlets within the Local Cavity should provide very weak absorption sites for
the more distant soft X-ray emission, thus not interfering with the X-ray spectroscopic and shadowing studies discussed previously.
However, only a very few cold (T $<$ 1000~K) and dense gas clouds and no molecular clouds, which can more efficiently absorb
soft X-rays, are known to exist within the Local Cavity \citep{lall03}.

\subsection{Transition Temperature Gas:  \oxysix}

One of the earliest pieces of observational evidence for a hot and highly ionized local plasma came from the detection of the \oxysix\ ion seen in the UV absorption spectra of nearby ($d < 150$~pc) B-type stars by the \copernicus\ satellite \citep{jenk74,jenk78}. These findings have been substantiated by the recent $\it FUSE$ surveys of \oxysix\ absorption detected towards hot white dwarf stars within 200pc of the Sun \citep{oeg05, sav06}. We note that \oxysix\ (1032\AA) is a tracer of transition-temperature gas at $\sim300,000$~K, and because of its quick cooling rate, theoretical production models require a source of hotter gas for its formation and continual replenishement.   The \oxysix\ could exist in a transition zone between the hot million K Local Hot Bubble  plasma and cooler gas clouds. The prevailing interpretation of these data is that the \oxysix\ resides on the periphery of the Local Cavity and/or on evaporating interfaces located between the (numerous) partially ionized cloudlets which are embedded within the Local Hot Bubble. The FUSE  \oxysix\ detections were only made on local sight-lines with distances $> 35$~pc and with typical values of column density of $\log N(OVI) \sim 13.0$ cm$^{-2}$.

However, recent re-examinations of \oxysix\ data forward evidence that argues against the presence of this ion being located within the rarefied confines of the Local Hot Bubble, whose contours have
been discussed in Section 2.2.  Firstly, for the 5 B-stars with (Hipparcos) distances $< 75$~pc  (i.e.
the nominal distance to the edge of the Local Cavity in most directions) the \copernicus\ data reveal only one (albeit tentative) detection of interstellar \oxysix\ absorption \citep{jenk78, bow08}. We note that there is a large variation between the (one) measured \oxysix\ column density and the (four) upper limit values for these very local sight-lines, which is unlikely to be due to any \oxysix\ absorption that would exist with a million K plasma. Instead, this effect has been attributed to a variation in the character of the transition temperature zones surrounding individual clouds.

A new detailed analysis of the radial velocities of interstellar and (when present) stellar absorption lines in a sample of $\sim100$ 
absorption spectra recorded by $\it FUSE$ for hot white dwarfs within 400pc of the Sun has recently been made available \citep{bar08,bar09}. For
2/3 of the sight-lines having significant OVI features  the velocity of the OVI
absorption is found to lie within the velocity uncertainty (i.e. 14 km s$^{-1}$) of features
that Barstow et al deemed to be
photospheric in origin, and thus an interstellar origin cannot be confirmed. In most of the remaining cases the sight-lines with confirmed interstellar OVI absorption
extend to or just beyond the neutral boundary of the Local Cavity and are at high galactic latitudes. Similar findings for the location of local \oxysix\ absorption, but
at lower latitudes, have also been made using a small sample of B-type stars located $>$ 80pc 
from the Sun \citep{welsh08}.

Since the local interstellar cloud is widely thought to be (at least partially) surrounded by a hot Local Bubble plasma, it could therefore be a prime evaporative transition region for the production of high ions.  Several authors have therefore constructed theoretical models that provide predictions for the numbers of associated \oxysix\ ions, which if spread over most of the cloudÕs surface should be detected in the absorption spectra recorded along most more distant sight-lines  
\citep{slav89, bork90, slav02, cox03}. Most models predict absorption column densities in the $\log N(OVI) = 12.5$ to 13.0 cm$^{-2}$ range, which is marginally consistent with what is detected by the $\it FUSE$ \oxysix\ surveys along local sight-lines. Since one might expect to encounter more than one conductive cloud interface over the $> 40$~pc sight-lines
 in which $\it FUSE$ has detected \oxysix, then the observations suggest that something may be reducing the effectiveness of thermal conduction in these \oxysix-producing interfaces. 
Jenkins \citep{jenk08} has proposed that local clouds could mutually shield each other from
 the effects of a hot Local Bubble plasma, such that a conduction front suitable for the formation
 of OVI only occurs at the very periphery of the cloud grouping. 
Alternately, others have have shown the importance of magnetic fields in their ability to inhibit conduction \citep{slav89,cox03}. Although the strength of the magnetic field in the case of the local interstellar cloud has been estimated to be of a low value of $\sim2\ \mu$G \citep{wood07}, if the magnetic field is tangled then this could account for the low values of \oxysix\ ion production along certain sight-lines.

\subsection{Size and Shape of the Local Hot Bubble}

The approximate size and shape of the emitting region of the local hot plasma
has been estimated under the assumption that the emissivity per volume is constant throughout the Local Hot Bubble \citep{snow90,snow98}.   
This assumption implies that the Local Hot Bubble's surface brightness in any given direction
is linearly proportional to its radius.    
The actual distance scaling was determined from 
observations aimed toward the shadowing molecular cloud MBM~12, 
which appeared to be placed within or just beyond the edge of the Local Hot Bubble.
The initial calculations assumed 
a distance of $\sim$ 65pc to this molecular cloud.
However, more recent estimates of the distance to
MBM~12 now place it at
a distance $>$ 200pc \citep{luhm01, chauv02}, such that the main
absorber of the more distant emission in this direction is instead provided by the neutral gas wall to the Local Cavity which is located at
a distance of $\sim$ 100pc. Thus, the revised radius of the hot bubble cavity should be about 1.5 times 
greater than the 50~pc to 100~pc range found previously.

Both the (previous and revised) estimated shape and size of the Local Hot Bubble derived from
X-ray observations do not match the detailed contours of the Local Cavity derived by the NaI 
mapping method and shown in Figures 2 and 3. 
If the Local Cavity is uniformly
filled with hot emitting plasma then the significant extension of the 
cavity to a distance of
$\sim$ 200pc in the direction of the star $\beta$ CMa would be expected to show an enhanced level
of soft X-ray background emission, which is not seen in the data \citep{snow98}.
However, there is  at
least some concordance that the first 50pc in all directions it contains very little 
cold and dense neutral gas (although there are several partially ionized clouds
within this volume).
Furthermore, the $\it ROSAT$ soft X-ray maps find that the Local Hot Bubble is unusually bright in 
two high 
latitude regions
that are offset by $\sim$ 45$^{\
m{o}}$ from the north and south poles \citep{snow98}.
These directions have been found to be well aligned with the openings
of the
Local Chimney into the halo \citep{welsh99}, as shown in Figure 3. Additionally,  it is well-known that there is a dipole effect in
the distribution the soft X-ray background emission at mid-plane latitudes, with the emission being
warmer, of greater intensity and with a larger spatial coverage in directions towards the
galactic center than sight-lines towards the galactic anti-center
\citep{snow00}. The present model of the Local Hot Bubble provides no substantial explanation
for this effect.
However, this effect can be explained through inspection
of the NaI density map shown in Figure 2, which clearly shows that there are many more gaps in the
neutral boundary of the Local Cavity at mid-plane galactic latitudes
between 190$^{\circ}$  $<$ $\it l$ $<$ 330$^{\circ}$
 (that could potentially
allow the passage of more distant generated soft X-rays) than in the opposite galactic direction.

Finally, we note that an alternative view has been forwarded in which the Local Hot Bubble is smaller
than the Local Cavity, leaving a
space  in the interstellar medium between the inner hot gas and the  outer dense gas of the cavity walls.   The properties of the 
material within this space are unknown, and how such a situation could have arisen or be sustained
has not been adequately explained.

\subsection{Thermal Pressure of the Hot Gas}

The derived pressure of the local hot gas has been a longstanding
 puzzle, together with the global problem of the apparent pressure imbalance
  between galactic disk gas and that of the overlying galactic halo \citep{cox87}.
 Supporting the halo requires a gas pressure of $P/k$ = 7,000 to 10,000~cm$^{-3}$~K \citep{shel07},
 suggesting that the Local Hot Bubble should have a similar pressure.
More recent local pressure estimates of $P/k \sim 15,000$~cm$^{-3}$~K have been derived from soft X-ray and EUV shadowing observations towards several nearby cold interstellar clouds \citep{snow93,kuntz97,burr98,berg98}.
In contrast, measurements of thermal pressures of several 
diffuse gas clouds residing inside the Local Hot Bubble result in far lower pressure values \citep{jenk02}. 
There clearly seems to be a pressure imbalance between the hot and colder gas within the LISM . In particular such an imbalance would tend to crush the cool clouds residing within the Local Hot Bubble and also then to try and expand the size of Local Cavity, neither of these effects has been observed.

However, this apparent local pressure anomaly could be accounted for in that 
(i) the previous distance estimates to the cold clouds used in these shadow 
experiments have now been revised (see Lallement et al 2003) such that a 
lower value of $P/k$ for the hot gas pressure is now obtained
(i.e.  $P/k \sim$ 12,250 cm$^{-3}$~K due to the revision of the 
clear pathlength in the MBM~12 direction from 65~pc to 100~pc, 
the distance to the Local Cavity boundary), and/or 
(ii) any remaining pressure difference could be accounted for by interstellar magnetic and/or turbulent gas pressure.  However, even with conservative revisions to these data, the spectre of a pressure imbalance still remains. A more thorough discussion of the issues that arise from this apparent
pressure imbalance can be found in Jenkins \citep{jenk08}.

\subsection{Kinematics of the Local Cavity Gas}

Although the main physical properties of the local hot gas have been determined from soft X-ray background observations, the limited spectral resolution of these emission-line data preclude a detailed kinematic assessment of gas velocity. However the motions of lower temperature gas residing within the Local Cavity can be derived from the absorption profiles of UV and optical lines observed at high spectral resolution along sight-lines towards nearby stars \citep{red08,sav06,craw98}.

For the case of the numerous partially ionized gas clouds of temperature $\sim 10^4$~K that are known to reside within $\sim50$~pc of the Sun, their velocity vectors are roughly parallel with a general flow originating in the Scorpio-Centaurus association. The velocity amplitudes encompass a wide range of values between clouds, suggesting that interstellar shocks may be prevalent. The turbulent velocity in some of these clouds is high, suggesting the presence or recent existence of shocks, which could result in the return of metals to the gas phase through the shock dissipation of dust grains \citep{red08}.

For higher temperature (T $\sim$10$^5$~K) gas located within the Local Cavity, most of the local \oxysix\ absorption velocities correlate well with the lower temperature ion velocities \citep{sav06}. This is to be expected if \oxysix\ is formed in a condensing interface between cool and hot Local Cavity gas. There are also some sight-lines in which the \oxysix\ absorption velocity is displaced to positive velocities from the lower temperature gas, which suggests in these cases that the \oxysix\ may
be tracing an evaporative flow from interface regions.

Measurements of the absorption velocity of the NaI neutral gas associated with the nominal boundary to the Local Cavity do not suggest that the Local Cavity wall is expanding or contracting with any significant gross motions (Lallement, private communication).  This would be expected if the Local Cavity is an old SNR whose expansion has been significantly slowed down by its interaction with the surrounding dense ISM. Also, if the Local Cavity is collapsing (such as being due to a marked pressure imbalance) then it would collapse at the sound speed, which is not observed.

Finally we note that there are two major dynamical movements of gas inflow into the Local Cavity region. The first, as mentioned earlier, is a flow of gas clouds into the Local Cavity from the adjacent Loop I superbubble (i.e. the Sco-Cen association). This inflow has been traced using both low (CaII, SII, FeII) and high ionization (CIV, SiIV, NV) species, all of which possess significant negative velocity motions for sight-lines of distance $> 90$~pc in the general direction of l = 330$^{\circ}$, b = +18$^{\circ}$ \citep{welsh05}. Secondly, for galactic directions close to the axis of the Local Chimney that extend into the lower halo, interstellar gas clouds with inflow velocities in the -20 to -60 km/s range have been detected along sight-lines spanning both open-ends of the Local Chimney which appear to be falling towards the galactic disk \citep{welsh04,craw02}. 

\subsection{Origin and Age of the Local Hot Bubble and the Local Cavity}

The history and age of the Local Hot Bubble are much debated. It is widely believed that the Local Hot Bubble and Local Cavity are the result of one or more supernova (SN) explosions \citep{cox82}, but in an alternate scenario this superbubble picture is disputed, and it is instead the rarefied Local Cavity that is thought to be caused by star formation epochs in the Scorpius-Centaurus OB association as regulated by the nearby spiral arm configuration \citep{boch87,frisch95,fuchs06,maiz01}. Other suggested Local Cavity formation scenarios involve an association with the young and massive stars of the  Gould Belt \citep{gren00} or the passage of a high velocity cloud through the galactic disk \citep{comer94}.  The present day morphology of the Local Cavity seems to be defined by the `voidÕ within the ISM created by the outer shells of nearby overlapping superbubble structures \citep{welsh94, frisch95}.

Recent modeling of the joint evolution of the Loop I and Local Hot Bubble superbubbles involve $\sim20$ SNe occurring in the moving group of OB stars of the Sco-Cen cluster which passed through the present
 day local cavity \citep{breit06}. In such a scenario the formation age of the Local Hot Bubble is constrained to $\sim14.5$~Myr, with the last re-heating occurring $\sim0.5$~Myr ago. In a similar supernova driven formation model the last re-heating episode
 is placed at $\sim3$ million years ago \citep{smith01}, based on the excess
 abundance of $^{60}$Fe found in the Earth's deep ocean crust \citep{fields05}.

\subsection{Solar Wind Charge Exchange (SWCX)}

At the Local Bubble conference in 1997, Freyberg \citep{frey98} and Cox \citep{cox98} separately recognized that the then-unknown X-ray production mechanism which made Comet Hyakutake surprisingly X-ray bright may operate elsewhere in the solar system. They warned astronomers that these X-rays could pose a significant problem for interpretations of the soft X-ray emission from the Local Hot Bubble. It was soon realized that the source of such cometary X-rays was the deexcitation of solar wind ions following charge exchange with neutral material \citep{crav97,crav00}. Very highly ionized and fast moving solar wind ions have a large cross section for interacting with neutral material in cometary and planetary atmospheres, as well as neutral material that has been swept into the heliosphere. During the interaction, called solar wind charge exchange (sometimes referred to as SWCX), an electron from the neutral material is transferred into a highly excited state in the solar wind ion followed by a decay that produces one or more high energy X-ray photons \citep{warg08}.

Coronal mass ejections, which enhance the solar wind density in localized regions for  hours to days, are probably responsible for the long term enhancements seen in the \rosat\ data and nominally removed from the \rosat\ All Sky Survey \citep{snow97,snow98}. The SunÕs slow and fast winds, which are less sporadic but vary in relative contribution during the 12 year solar cycle, create a slowly varying component which has not been removed from the \rosat\ All Sky Survey. Since solar wind charge exchange X-rays are produced locally, they also behave as a local component in all shadowing observations. This signal level has generally been attributed to emission associated with the Local Hot Bubble. Lallement \citep{lall04}, Cravens et al \citep{crav01} and Robertson \& Cravens \citep{rober03} have assessed the potential level of contamination to the \rosat\ All Sky Survey by solar wind charge exchange in the slow and fast winds.  Robertson $\&$ Cravens \citep{rober03}) found that 0.1 to 1.0 keV X-rays produced in the heliosphere by the steady slow and fast winds could account for $\sim$ 1/2 of the soft X-rays in the Galactic plane and a smaller fraction of the high latitude locally-made soft X-rays. Similarly, Lallement \citep{lall04} calculated the steady, heliospheric solar wind charge exchange X-ray intensity in the 1/4 keV band, the band in which the local X-ray emission is most prominent, finding that solar wind charge exchange could explain most of the 1/4 keV intensity that was previously attributed to emission from the Local Hot Bubble. Both sets of estimates are uncertain, due to uncertainties in the neutral ion density, solar wind density, interaction cross section, and solar wind charge exchange spectrum, making precise estimates difficult. 
Even when the upper limits are used in these global predictions, 
the level of predicted solar wind charge exchange X-ray
emission cannot explain the excess of local X-rays observed at high latitudes.
In addition, the maps of Roberston $\&$ Cravens \citep{rober03} show significant 
structure at low latitudes.   The \rosat\ maps do not show corresponding
dim regions, suggesting that the Local Hot Bubble's 
X-ray emission has compensated for inhomogeneities in the solar wind charge exchanges.   
However, an alternate conclusion is that
the solar wind charge exchange X-rays are more evenly distributed than in the
Robertson $\&$ Cravens maps.

Koutroumpa et al. \citep{kout06,kout07,kout08} have made more detailed calculations of the solar wind charge exchange spectrum, finding it to be somewhat harder than that attributable to the nominal Local Hot Bubble. As a result, they conclude that solar wind charge exchange can explain all of the local 3/4 keV emission found from the \rosat\ All Sky Survey and all of the local \oxyseven\ 570 eV and \oxyeight\ 650 eV line emission seen in pointed observations by \chandra, \xmm, and \suzaku.  More recent calculations by
Koutroumpa et al. \citep{kout09} that compare
the calculated solar wind charge exchange spectra
with soft X-ray shadow fields observed with \rosat\ reveal that the solar wind charge exchange intensity is approximately independent of latitude.
The intensity of the solar wind charge exchange emission in the \rosat\ 1/4~keV band is, 
on average,
$\sim 332 \times 10^{-6}$ counts s$^{-1}$ arcmin$^{-2}$, which is of the same order as the soft X-ray background intensity that was previously attributed to emission from the Local Hot Bubble at low galactic latitudes. However in high latitude
regions ($b > \pm 40^{\rm{o}}$) the calculated levels of solar wind charge exchange emission cannot explain the high
levels of 1/4 keV emission observed by \rosat\ and thus a hot emitting gas in the galactic halo
is required.

The calculations of Koutroumpa et al  suggest that the solar wind charge exchange spectrum may be harder than previously thought.  This notion agrees with the observations of the soft X-ray background spectrum \citep{sand01,mccam02}. However, we note that these latter data were collected from a large area of the sky centered close to the axis of the Local Chimney at high northern galactic latitudes and thus presumably sampled significant amounts of (distant) emission from the hot and ionized
million degree K  halo. Both sets of the soft X-ray background measurements cannot explain all of the observed 1/4 keV emission and even less of the B-band emission recorded by the Wisconsin surveys, in addition to being in conflict with the EUV spectral data. Thus, the Local Hot Bubble spectrum is required to be softer then previously thought, and the soft X-ray emission shortfall could be
 filled by a warm (T $<$10$^{6}$K) emitting gas within the Local Cavity.

\section{Ramifications for the `Accepted Local Hot Bubble Model'}

It is clear from the preceding sections that several important problems 
plague the traditional model of the Local Hot Bubble. 
The most important of these is the contamination by solar wind charge exchange X-rays.   
The predicted severity of the 1/4 and 3/4 keV X-ray
 contamination by this heliospherically generated emission for low latitude sight-lines ranges from
  $\sim1/2$ in the model of Robertson \& Cravens \citep{rober03}
to approximately 100$\%$ from the calculations of 
Koutroumpa \citep{kout09}. 
Thus, new models of
the local region range from those that are $\sim1/2$ as bright in X-rays
as the traditional Local Hot Bubble models to those that 
have no local hot gas in the Galactic plane.   In either case, the
model must include X-ray producing gas located at high latitudes
because solar wind charge exchange models cannot explain all of the observed high latitude
X-ray emission.
In this subsection, we consider the first case in which the
Local Hot Bubble is half as bright in the plane as previously
believed.   
In the following section, we consider the other
extreme, that in which gas in the local ISM produces none of the soft X-rays seen
at low latitudes.

Firstly, if emission from a Local Hot Bubble is diminished relative to traditional
models, it is diminished at all latitudes
by a similar amount ($\sim 330 \times 10^{-6}$ counts s$^{-1}$
arcsec$^{-2}$ in the \rosat\ 1/4 keV band) according
to Koutroumpa et al. \citep{kout08}.
Subtracting a constant intensity from the observed anisotropic
distribution of locally produced X-rays leaves a more extremely
anisotropic distribution of X-rays which can then be attributed to
the Local Hot Bubble.
If we make the standard assumption that the X-ray emissivity
is constant throughout this emitting region, 
then the Local Hot Bubble must be extremely distorted, with 
strong lobes in the northern and southern hemispheres and a tight waist
in the Galactic plane.
Secondly, 
the model solar wind charge exchange spectra are harder than the
observed spectrum.
We conclude this because
the solar wind charge exchange predictions of Koutroumpa et al. account
for a greater fraction of the observed X-rays in the 1/4 keV 
band than in the Wisconsin B band (130 to 188 eV).
In order to compensate for the hardness of the solar wind charge exchange spectrum,
the Local Hot Bubble spectrum must be softer 
than previously believed.   Thus, the local emitting plasma would need to be cooler than
previously believed.  

Any revisions to the X-ray brightness and temperature require that 
the electron density be recalculated.
If the temperature and path length are unaffected, then
a reduction in the Local Hot Bubble emission intensity by a factor of 2 would imply a 
reduction in the electron density by a factor of 1/$\sqrt 2$,
resulting in $n_e \sim0.005$~cm$^{-3}$.
This serves as a reasonable starting point.
However, 
once the solar wind charge exchange contamination is better understood, this estimate
can be improved upon by an improved derivation of the temperature of
the Local Hot Bubble plasma (and thus the emissivity) and the intensity
of Local Hot Bubble X-rays in the direction of MBM12.
Because the electron density and temperature are less than previously believed,
the thermal pressure must also be less.    The
reduction in the electron density by 1/$\sqrt 2$ alone, brings the
estimated thermal pressure down from 12,250~K~cm$^{-3}$ 
to $\sim 8,700$~K~cm$^{-3}$.   Any revision to the local
plasma temperature will
will lower the gas pressure further.    

These revisions reduce or resolve some of the problems posed by observations.
Firstly, 
it may now be possible for the hot gas in the Local Cavity to be in 
pressure balance
with the embedded warm clouds, especially if the expected
downward revisions to the Local Hot Bubble's thermal pressure are severe and if
the clouds 
are somewhat supported by non-thermal pressures such as magnetic pressure
and turbulence.
Furthermore, the reduction in the Local Hot Bubble X-ray intensity may help to 
explain the low fluxes of the \ironnine, \ironten, and 
\ironeleven\ photons observed by in ultra-soft X-ray spectra
\citep{mccam02,hur05}.  The precise
estimates of the expected \ironnine, \ironten, and \ironeleven\ 
intensities, however,
depend on the Local Hot Bubble's  (newer lower) temperature, which is not yet known.

These revisions also re-open existing inquiries.   The heating mechanism
for the hot gas must be re-evaluated now that the gas is found to be
dimmer and cooler, and the spatial distribution of the hot gas must also be
reconsidered.    The previously estimated shape of the emitting cavity, which was
determined from the 1/4 keV X-ray intensity and the assumption that the 
intensity scaled with the pathlength, disagreed with
the shape of the Local Cavity, but such disagreement may evaporate
once the spatial distribution of Local Hot Bubble X-ray emission is better known.
In addition, given that the bubble's X-ray intensity is less than previously
thought, the source and history of the Local Hot Bubble must be reworked. A single
or multiple supernova may have heated the gas, but such events were probably fewer
and possibly happened further back in the past than currently believed.

As previously noted, some of the observed physical parameters of the LHB appear to be
consistent with the characteristics of a fossil SNR. The Local Cavity is not expanding
at the expense of its surroundings,  just as fossil remnant cavities are no
longer driven by  the over-pressure of expanding gas within them.
However, the hot gas once contained within the Local Cavity may have
escaped through the Local Chimney into the
galactic halo, or it may have cooled due to turbulent mixing or
conduction with entrained gas or gas close to the galactic plane.
The Local Cavity structure may now be gently photo-ablating in the
residual UV field.   Shelton \citep{shelt98} has discussed
the possibility that the Local Cavity may be an aging SNR.
In her model, she notes that once a cool expansion shell forms
any OVI should form within and at the periphery of
the hot bubble, which is moving far
slower than the SN shock front.  Once the SNR shell forms,
the soft X-ray emission (from the remnant hot gas) ceases to be
edge-brightened and its emission diminishes by
an order of magnitude. In addition, she finds that the X-ray
productivity of older SNRs depends critically on the ambient
non-thermal pressure, thus allowing the possibility that an
ancient SNR in a diffuse environment may possess a higher
pressure than previously thought.  Because such a model
does not explain the observed
local spatial distribution of OVI which appears to be mostly
formed at
the periphery of the cavity as opposed to being
distributed within the cavity itself, we suggest that most
of the observed O VI is due to the interface between the cavity
gas and the hot halo gas.
In addition, without an additional
source of heating, old remnants tend
to collapse such that the
material around a fossil SNR will have moved back to fill in the low
pressure space, and there will not
be an observable cavity.  

We now turn to a speculative model of the Local Cavity in which solar wind charge exchange
explains all of the observed soft X-ray emission 
in the Galactic plane, i.e. the local interstellar medium is devoid of hot gas in the plane. In addition,
the cavity is filled with warm (20,000K) photo-ionized gas,  some of which may
have entered the aging cavity from the Sco-Cen bubble. Although such a model represents
a radically different view, it goes a long way in resolving many of the outstanding
issues that currently bedevil the generally accepted model of a Local Hot Bubble.

\section{The `Hot Top Model'}

The following is a possible scenario for the physical state of the Local Cavity. It is not based on a rigorous theoretical framework but is, instead,  driven directly by recent observational evidence.
Although it is far from complete, it attempts to accommodate the ramifications of the recent observations of solar wind charge exchange, OVI and  the EUV background emission. Although referred to here as a `model', it is really a descriptive
framework designed to stimulate theorists into re-assessing the currently accepted models of a hot Local Bubble.
In addition, it has testable predictions that future observations in the UV and/or soft X-ray regimes will be able to
confirm or refute.
Bearing these caveats in mind, we assume the following:

The size and shape of the Local Cavity is defined by the neutral contours derived from NaI absorption measurements (Lallement et al 2003), such that the absorbing wall that surrounds most of the Local Cavity can be categorized by $\log N(HI) > 19.3$~cm$^{-2}$. A possible new picture of the Local Cavity is one in which it is mostly filled with a very diffuse and photoionized and/or collisionally ionized gas of temperature $\sim20,000$~K that is in rough pressure equilibrium ($P/k \sim 3000$~cm$^{-3}$) with the numerous partially ionized and warm ($T \sim 7000$~K) small cloudlets that are known to exist within the LISM. 
The average electron density of the ionized Local Cavity `filler gas' is assumed to be 
$n_e \sim 0.04$~cm$^{-3}$ (derived from pressure equilibrium with the partially
 ionized LISM cloudlets)
and there is also a very low average neutral gas density of 
 $n(H) < 0.01$~cm$^{-3}$ within the Local Cavity \citep{dup98,welsh94}.
We note that there may also be a contribution
from magnetic pressure such that the total pressure of both the local interstellar cloud and `filler gas' may be higher
than we have assumed. \footnote{Although we use the term `filler' gas, in reality the photo-ionized regions will
be the rarefied ionized envelopes surrounding cooler interstellar cloud cores. Since the
Local Cavity has numerous cloudlets of this type, these outer cloud envelopes will probably merge
into a photo-ionized medium that can best be described as a `filler' gas.}

Our model also requires that
the major source of local non-solar wind charge exchange generated soft X-rays occurs at high galactic latitudes ($b > +/-35^{\rm{o}}$), which is probably due to emission from a hot and ionized gas viewed
through the openings of the Local Chimney in to the overlying halo.   Also there is a highly absorbed component generated in surrounding hot (X-ray emitting) superbubble cavities that abut the Local Cavity within the thick galactic disk. A simple schematic of the model is shown in Figure 3.

Our assumption that the Local Cavity is mostly filled with warm and photoionized gas 
is based on observations of the \argonone, \nittwo, \carthree\  and \silthree\ ions that have been widely observed throughout the local ISM \citep{lehn03} and by the recognition that the shortage of neutral argon in nearby clouds
can be better explained by photoionization rather than collisional ionization \citep{sof98}.  In addition, photo-ionization
 conditions are observed to vary considerably throughout the LISM \citep{lehn03}. For example, Wolff et al. \citep{wolff99} have found evidence for an ionization gradient in the local gas with an ionization increase along the general direction to the Canis Major extension to the Local Cavity. 
 
Most back-of-the-evelope calculations assume that the plasma is in collisional ionization equilibrium throughout
the Local Cavity, although there is no real consensus on this matter and such models do not fit the observed soft
X-ray background spectrum \citep{sand01,breit01}. Bruhweiler $\&$ Cheng \citep{bruh88} and Vallerga \citep{vall98} have
taken a different approach by considering the photoionizing effects of nearby hot stars. Photoionization from
the local EUV radiation field (which is dominated by the EUV flux from the two B stars $\epsilon$ and $\beta$ CMa, together with a few nearby hot white dwarfs) can account for the high interstellar gas electron density measured towards nearby
stars, although the spectrum is too soft to explain the apparent over-ionization of helium with respect to
hydrogen measured in the local cloud \citep{vals98}. 
Several authors have invoked additional (non-equilibrium) sources of ionization to explain this anomaly \citep{slav02}. In one model, the shocks that resulted from
a local supernova event that may have occurred $\sim$ 3.6 million years ago are the main cause
of  the time-dependent ionization at the Sun, although most of the photons with wavelengths $>$ 500\AA\ are due to the
stellar spectra \citep{lyu96}. Thus, the supposed shock ionization event, from which the gas has yet to return to its
equilibrium state, is supplemented by the present stellar radiation field. The ionization rate for this
is $\Gamma$(HI) = 1.1 x 10$^{-15}$s$^{-1}$ \citep{vall95}.  We note the long recombination time of hydrogen in the ISM, which for 10$\%$ ionization and n(HI) $\sim$ 0.1 cm$^{-3}$ is about 10$^{7}$ years \citep{chas86}. The timespan since the supposed supernova explosion also agrees, roughly, with the 
 data gained from the ferromanganese ocean layer crust  of the Earth, for which a local supernova event is required to have happened $\sim$ 3 Myrs ago \citep{knie04,fields05}. Thus, although we have invoked photo- (and some collisional) ionization  in our new model, the long time-scales for H recombination and the current availability of sufficiently strong stellar photo-ionization sources seems adequate to sustain both the ionization and temperature of our theorized Local Cavity filler gas.

From the various solar wind charge exchange calculations of Koutroumpa et al, Lallement, and Robertson $\&$
Cravens, we know that there may be far less X-ray emitting interstellar gas 
in the local region than previously expected.    
As noted in Section 2.8, simulations of the contribution of solar wind charge exchange spectra in 
\rosat\ R1 and R2 observations of shadow field sight-lines by Koutroumpa et al \citep{kout08,kout09} indicate that the level of solar wind charge exchange flux is not strongly latitude dependent, but can account for almost all of the observed R1 and R2 emission at low galactic latitudes
(i.e. within the Local Cavity). However,
an unabsorbed 1/4 keV soft X-ray background component is seen to be well correlated with the absolute galactic latitude for galactic latitudes of b $>$ $\pm$35$^{\circ}$.
Thus, a high latitude interstellar component exists and the spatial distribution of its emission
seems directionally correlated with sight-lines viewed
through both openings of the Local Chimney into the lower
galactic halo \citep{welsh99}.

We therefore conjecture that at high latitudes the observed soft X-ray background emission (which is thought to arise in a hot and highly ionized halo) is only weakly absorbed when viewed through the low \hone\ column
sight-lines that are contained within the `openings' of the Local Chimney, and thus
the soft X-ray background is detected with a greater intensity in these directions.  For sight-lines into the halo that do not pass through these low opacity openings, the soft X-ray background emission signal is highly absorbed by
the higher \hone\ density  gas in the galactic plane.
This is in accord with the  \rosat\ shadow observations which
argued against a single emission component from a hot halo of large scale-height \citep{snow00},
but instead required a low scale-height variable
(in both temperature and intensity) emission component,
which could also be considered as the superposition of
multiple, spatially distinct emitting components.
Additionally, the ends of the Local Cavity near the Chimney openings may be filled with
hot gas that has flowed into it from above.

We know that there is anomalous structure in the direction of the northern
chimney opening because of polarization data.
Mathewson $\&$ Ford \citep{math70} have measured the polarization to nearby stars and found that for sight-lines $< 50$~pc the E-vectors have low polarization values and are also well ordered with no apparent major disturbances in the local magnetic field, as also concluded by Tinbergen \citep{tin82}. However, for sight-lines in the 50 to 100~pc range there is a clear change in both the magnitude and direction of the E-vectors which rise up from the galactic plane to high galactic latitudes spanning the galactic longitude range $\ell = 300$ to +30$^{\rm{o}}$. Additionally, using radio polarization data, 
Wolleben \citep{woll07} has formulated a new model for the Loop I radio feature that 
involves two X-ray emitting shells. The region of overlap of these two 
expanding shells of Loop I matches most of the disturbed region of E-vectors 
in the 50 -100pc range noted by Mathewson $\&$ Ford \citep{math70}. A similarly disturbed region
is also predicted by Wolleben at the southern opening of the Local Chimney.

The idea for our new picture of the Local Cavity was largely driven by an interpretation of the results from a recent re-analysis of  \oxysix\ absorption profiles detected by the FUSE
satellite towards local hot white dwarfs \citep{bar08,bar09}. Apart from a few instances,
the new Barstow et al. results largely agree with and extend upon previous measurements which
contained a smaller sample of hot white dwarf targets
whose spectra were reduced using an earlier version of the FUSE data  processing pipeline \citep{oeg05,sav06}.
In Figure 4 we show the spatial distribution of the 1/4 keV R1 and R2 soft X-ray background emission \citep{snow98} together with the galactic positions of local (d $<$ 200pc) hot white dwarfs that have undisputed detections (and non-detections) of interstellar \oxysix\ absorption (i.e. those sightlines in which the velocity
difference between interstellar and photospheric OVI absorption is greater than the FUSE
spectral resolution). In addition we have similarly plotted OVI data from the 4 sight-lines within 120pc \citep{sav06} ) that are not part of the new OVI data set \citep{bar09}. Finally,
for completeness,  OVI detections and non-detections towards
B-type stars with distances $<$ 120pc 
are also included on the plot  \citep{bow08,welsh08} .

It is clear from Figure 4 that there is a remarkable spatial co-incidence between detections of local \oxysix\ absorption  (d $<$ 80pc) and the highest levels of soft X-ray background emission. This is to be expected if \oxysix\ absorption is formed in a transition region between a hot ($T \sim 10^6$~K) and cooler ($T \sim 20,000$~K) plasma. The
spatial distribution of \oxysix\ absorption shown in Figure 4 can be explained by our presently proposed model if
local \oxysix\ absorption ($d < 100$~pc) only occurs in only two environments: (i) at the
interfaces between the Local Cavity and the overlying galactic halo (at the `top' and `bottom' of the Local Chimney) and (ii) at the neutral walls between the Local Cavity
and nearby hot bubbles.
The former
distribution of \oxysix\ absorption is confirmed by the
new OVI results in which \oxysix\ is only detected at high
latitudes for sources with distance $>$ 60pc \citep{bar08,bar09}. 
The distribution of \oxysix\  at low latitudes is consistent with its formation
at the interfaces between distant  (d $>$ 80pc) bubble cavities and their surrounding
neutral gas shells \citep{bow08,welsh08}. 

We conjecture that the above model can also explain many of the problems associated with the widely held hot gas model of the Local Hot Bubble listed in Section 2. In particular, this new model provides the following explanations for many of the problems outlined previously.

\subsection{The Local Cavity Pressure Problem}  
There are now only two types of gas contained within the confines of the Local Cavity, i.e. a strongly photo-ionized ionized `filler gas' at $T \sim 20,000$~K and numerous partially ionized cloudlets of $T \sim 7000$~K, both of which may be in approximate pressure equilibrium. The million $^{\
m{o}}$K, soft X-ray emitting gas is now confined to high latitude sight-lines 
at distances of at least 50 - 100pc from the Sun. The hot, rarefied gas in the halo has a lower mass density than that 
in the disk, but it has a greater thermal pressure of $\sim$7 to 10 x 10$^{3}$ K cm$^{-3}$ for OVI-rich gas \citep{shel07}.
Ironically, in re-conceptualizing the local region, the thermal pressure has gone from being larger than that of the 
hot halo gas to being smaller. However, thermal pressures on the order of $\sim$ 3 x 10$^{3}$ K cm$^{-3}$
are common throughout the disk, making the discrepancy between disk and halo pressure a widespread problem. Also,
it is commonly assumed that other forms of pressure (including magnetic, cosmic ray and turbulent pressure) can
make significant contributions to the total pressure of the ISM. The magnetic field strength of the neutral wall
surrounding much of the Local Cavity in
the disk (estimated at $\sim$ 8 $\mu$G, Andersson $\&$ Potter \cite{anders06}, resulting in a
magnetic pressure of B$^{2}$/ 8$\pi$k of  $\sim$ 18,000 K cm$^{-3}$) demonstrates that it is possible for large
magnetic pressures to exist. A  magnetic field of this strength is very similar to that theorized for magnetic flux tubes
that were originally invoked to constrain the number of production sites for interstellar OVI in the local
ISM \citep{cox03}.

In addition, we note that 
global dynamic pressure equilibrium may not actually exist within the Local Cavity for several reasons. Firstly, the cavity is not fully pressure constrained by surrounding neutral interstellar gas walls and also there is clear evidence for the inflow of gas from both the adjacent Sco-Cen/Loop I cavity and from the lower galactic halo. Interestingly, it has been found that positive velocity OVI is mostly found at high latitudes in the northern galactic hemisphere, which has been interpreted as possibly being caused by an evaporative flow from a young interface between warm gas and a hot exterior (halo) medium \citep{sav06}.  This scenario is similar to our Hot Top model in which the evaporative outflows of OVI are formed at the interface between the warm photo-ionized filler gas of the Local Cavity and that of the hotter infalling exterior gas of the inner halo located at the two openings of the Local Chimney.

\subsection{Size and Shape of the Local Cavity} 
In the Hot Top Model, the size and shape of the rarefied Local Cavity 
are simply defined by the 
neutral gas walls (as traced by NaI absorption).   
In previous explanations for the Local Hot Bubble, either 
the hot gas was assumed to fill the Local Cavity and thus take
on its size and shape, or to fill a smaller region within the 
cavity such that the bubble's radius in any given direction
scaled with the observed local X-ray intensity.
The Hot Top Model easily resolves the difficulty with the first
version of the Local Hot Bubble model and its inability to explain why,
if the Local Cavity were filled with hot gas, the
$\beta$ CMa tunnel direction wasn't found be brighter in soft X-ray emission than 
neighboring directions.
This is no longer problematic if 
the tunnel is instead filled with lower temperature 
photo/collisionally ionized gas, as verified by several UV absorption 
studies of this region \citep{dup98,gry01}. In addition, our model removes
the somewhat contrived version of the Local Hot Bubble model which required a gap between the
soft X-ray emitting hot gas and the neutral gas walls of the Local Cavity.

\subsection{Origin and Age}
We note that interstellar structures similar to the Local Cavity and
the Local Chimney have been detected elsewhere in both our own and other galaxies \citep{mclur03}.
Therefore, the presence of such a low-density cavity in the ISM that is linked to an overlying galactic
halo is clearly not a rarity. Since our new model does not require the presence of significant amounts of local million degree K gas, then this suggests that either the Local Cavity is an old (fossil) SNR (T $>$ 10$^{6}$ years) in which any hot gas has long since cooled and/or recombined, or that the Local Cavity was formed by some other manner, perhaps that which created Gould's Belt \citep{lall08}.

The origin and evolution of local blow-outs from the galactic disk have been detailed using hydrodynamic
simulations \citep{breit06,avil07}.  If the
pressure of overlying halo gas is lower than the mean pressure in the disk, then interstellar bubbles (caused
by the action of either stellar winds or supernova events) could blow out of the disk in the polar direction. 
Such structures would cool rapidly in the equatorial regions and OVI could arise in the upward traveling shock,
and in the interfaces of the shocked cloudlets being ejected into the halo. We note that the numerous neutral and
partially ionized gas clouds
detected at both ends of the Local Chimney have mostly negative velocities, whereas the OVI bearing gas has
mostly positive velocities \citep{sav06}. 
Thus, the Hot Top Model is really an observational description of an aging SNR whose hot
gas has escaped and it is now filled with warm and
photo-ionized gas, which explains the observed local dearth of HI. This warm gas may have entered the aging cavity
through the adjacent Loop I interface, having being expelled as stellar winds from the Sco-Cen OB
association.

\subsection{Kinematics of the Local Cavity Gas} 
The motions of both ionized and partially ionized gas within the Local Cavity  (as measured by 
UV-visible absorption lines towards nearby stars) is governed by inflow 
from the adjacent interstellar bubbles that abutt the Local Cavity (i.e. Loop I, 
Loop III, Pleiades etc). No expansion is required (due to any pressure 
imbalance). The inflow of gas from the inner halo is seen only at 
$|z| > 100$~pc and the lack of any observed general inflow from this 
infalling gas into the very inner regions of Local Cavity is most probably due to 
such an effect being masked by the gas flow vectors from the adjacent 
interstellar cavities.

\subsection{Absorption Column Densities}
In order for \oxysix\ absorption to occur, an interface between hot and cooler gas is required. In the new model this can occur at locations within the Local Chimney where infalling highly ionized million degree halo gas meets the cooler photoionized filler gas of the Local Cavity.  The new \oxysix\ results support this scenario for sight-line distances $< 100$~pc \citep{bar08,bar09}.
For sight-lines in the galactic disk that extend beyond the confines of the Local Cavity, \oxysix\ absorption can occur at interfaces between more distant hot bubble cavities and their surrounding neutral gas shells \citep{welsh08,bow08}. Clearly more
distant (d $>$ 200pc) \oxysix\ absorption can also occur in more distant halo regions, as well as
in distant hot superbubble cavities throughout the galactic plane.

\subsection{Soft X-ray background and EUV Emission $\&$ Remaining Problems} 
Clearly the most radical proposition in our new model is the lack of a requirement for a ubiquitous million degree soft X-ray emitting gas contained within the confines of the Local Cavity,
as delineated by its neutral absorbing walls. Although the contribution of solar wind charge exchange generated emission may account for a large fraction of the observed soft X-ray background (especially at low galactic latitudes), some of the models still require some contribution from a million degree gas present at low latitudes. This contribution is thought to be small ($< 200 \times 10^{-6}$ \rosat\ counts s$^{-1}$ arcmin$^{-2}$) and we propose that if it exists then its intrinsic variation in intensity with (galactic) sight-line sampled has previously been masked by the far larger contribution to the total soft X-ray background signal generated by the foreground solar wind charge exchange emission. 
Therefore, if such hot gas does exist, then at present we know very little about its properties and previous models of
the Local Hot Bubble require substantial revisions.

Although the new model attempts to provide explanations for many of the observational characteristics
of soft X-ray, EUV, UV  and visible observations of the local interstellar gas, there is
one set of observations that cannot be accounted for. From sounding rocket data
gained over 25 years ago
numerous observers have found that the B/Be band ratio is essentially constant, irrespective
of the sight-line probed. However, present calculations of the solar wind charge exchange emission at these very
low energies suggests that very little emission arises within the B-band. 
We note that satellite observations of the diffuse EUV background \citep{jel95}
require limits for the plasma emission measure of a factor of 5-10 below the B-band emission measured over a temperature range from 10$^{5.7 -6.4}$ K. Also, the Be-band measurements by Cash, Malina $\&$ Stern \citep{cash76},
in agreement with the EUV satellite data, 
place a strict temperature upper limit of $\sim$ 3 x 10$^{5}$K to any emission from a local plasma. Such
conclusions are in direct conflict
with other measurements gained at these soft energies \citep{bloch86, juda91}.
Thus, the origin and physical nature
of this near-isottropic flux, in the context of our new model, still remains to be found. However, we
note that it
must be due to a local hot gas, but with a lower temperature than 1 million $^{\
m{o}}$ K. One
intriguing possibility is that such emission could arise in the 
terrestrial magnetosheath via charge exchange processes between heavy solar wind ions
and geocoronal neutrals \citep{robertson02}.

Clearly, new high spectral resolution observations of the background emission in
the 0.1 to 0.25 keV band will be required in order to solve this outstanding issue.
Such observations are non-trivial since they will require knowledge of the
separate contributions to the total emission signal from solar wind charge exchange and a presumed more
distant emitting hot gas. 

\section{Conclusion}

The interstellar region surrounding the Sun to a distance of $\sim$ 80pc in all directions
has historically been termed the Local Hot Bubble. Many of the physical properties of this gas,
as revealed by observations of the diffuse soft X-ray background (soft X-ray background) emission,
have long been described by emission from a million degree gas
in thermal equilibrium. In this paper we have presented a wide variety of observational
evidence gained over the last decade that would appear to contradict the expectations of 
a hot Local Bubble model. In particular, the recently realized non-negligible contribution
to the soft X-ray background signal from solar wind charge exchange X-rays generated within the heliosphere has required a critical re-thinking of the various physical components that constitute
the soft X-ray background spectrum.

We have provided arguments that would imply that, at the very least, the intensity of
the soft X-ray background emission arising solely from a local million degree hot plasma must be reduced by 
$\sim$ 50$\%$. We have outlined the ramifications of such a reduction in the
X-ray brightness
of a Local Hot Bubble and, although several observations that have seemed at odds
with the notion of an emitting Local Hot Bubble are now accounted for, 
the heating mechanism
for the hot plasma must be critically re-evaluated now that the gas is found to be
dimmer and cooler. In addition, the spatial distribution of the hot million degree gas
within the Local Cavity also requires reconsideration.

Finally we present a new model of the Local Cavity (i.e. the `Hot Top' model)
in  which solar wind charge exchange can explain all of the observed soft X-rays
in the Galactic Plane, and thus the local interstellar medium is devoid
of million degree plasma at mid-plane latitudes. Instead, the Local Cavity is
filled with a very diffuse and photoionized and/or collisionally ionized gas of temperature $\sim20,000$~K that is in rough pressure equilibrium ($P/k \sim 3000$~cm$^{-3}$) with the numerous partially ionized and warm ($T \sim 7000$~K) small cloudlets that
are present within the local ISM.
However, our new model does require
the presence of a million degree emitting gas at high galactic latitudes, such that
the associated soft X-ray emission is only
detectable when viewed along sight-lines that extend beyond both ends
of the Local Chimney feature into the lower galactic halo. The model also predicts
that locally, the OVI ion should be preferentially formed in interface regions between
warm photo-ionized gas and the hot million degree gas of the halo. This scenario
seems to be the case for OVI absorbers observed over sight-lines $<$ 80pc from
the Sun. The new model also provides an explanation for several observed parameters
such as local gas pressure, the observed `dipole' effect in maps of the soft X-ray background and
the spatial concordance between the highest levels of soft X-ray emission and the
spatial distribution of local OVI absorbers.

Although the Hot Top model can explain many observational
conundrums, it presently provides no explanation for observations of
the emission observed by sounding rockets in the B and Be-band X-rays. Such data are
also in contradiction with observations of the EUV diffuse background emission recorded by
satellite instrumentation.
Future high resolution spectral observations with an X-ray microcalorimeter at
ultra-soft X-ray wavelengths could provide important constraints on the (still)
disputed physical nature of interstellar gas within 100pc of the Sun.

\acknowledgments
BYW acknowledges financial support from NASA grant $\#$ NNX07AE33G.
RLS acknowledges financial support from NASA grant $\#$ NNH072DA001N.
Both authors would like to thank Don Cox, Steve Snowden, John Vallerga, Kip
Kuntz, Martin Barstow, Ed Jenkins and Rosine Lallement 
for many interesting discussions 
concerning this topic, most of which took place
while musing over a mug of fermented hops. We would also like to
thank both the editor and referee for their help in improving this paper.

\newpage

\begin{figure}
\center
{\includegraphics[height=5cm]{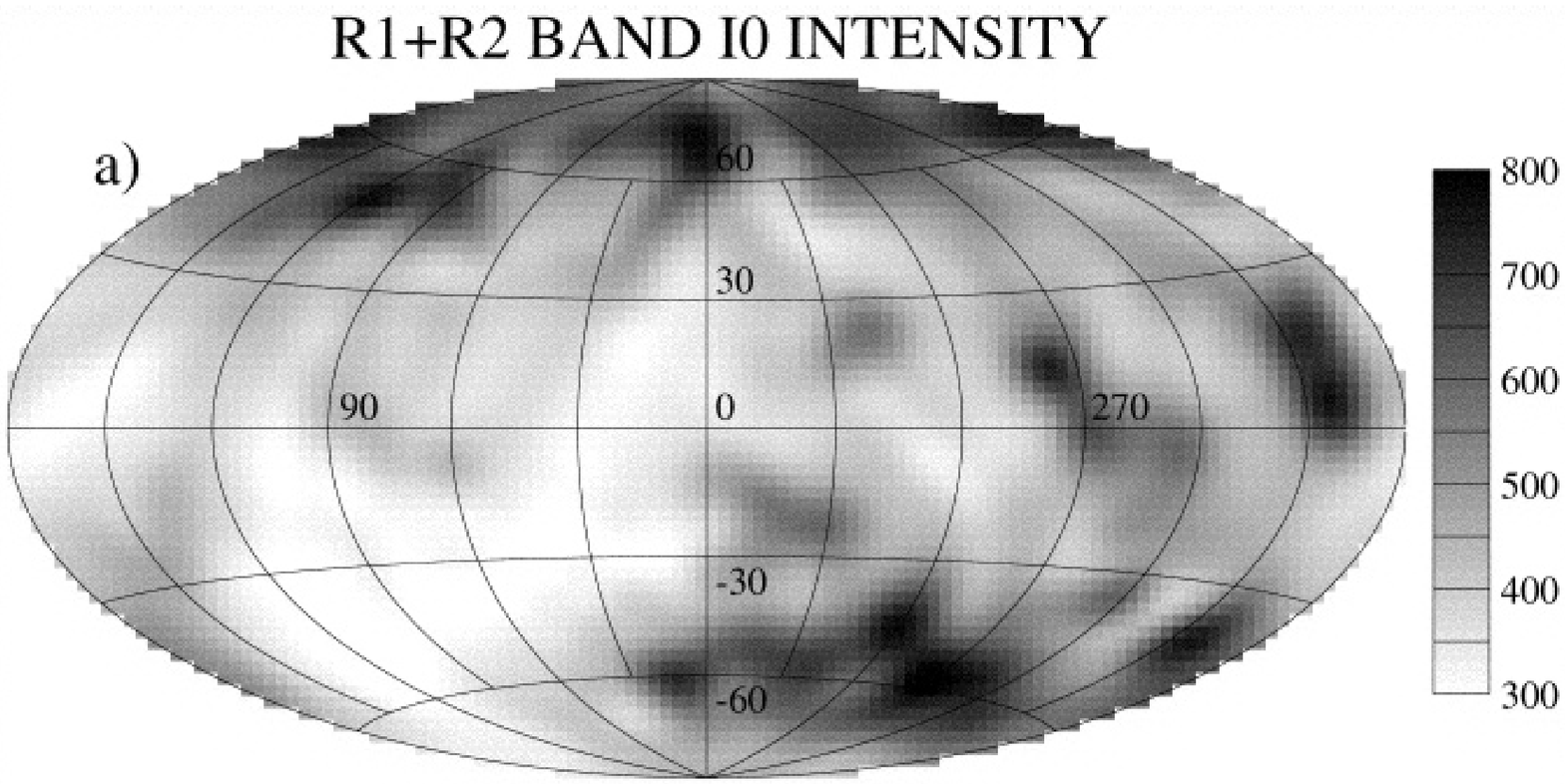}}
\caption{The galactic distribution of the summed R1 (0.11 - 0.284 keV) and R2 (0.14 - 0.284 keV) band emission as recorded by the $\it ROSAT$ all-sky survey \citep{snow98}. Note that the highest intensities are generally found at high galactic latitudes.}
\end{figure}

\newpage

\begin{figure}
\center
{\includegraphics[height=5.5cm]{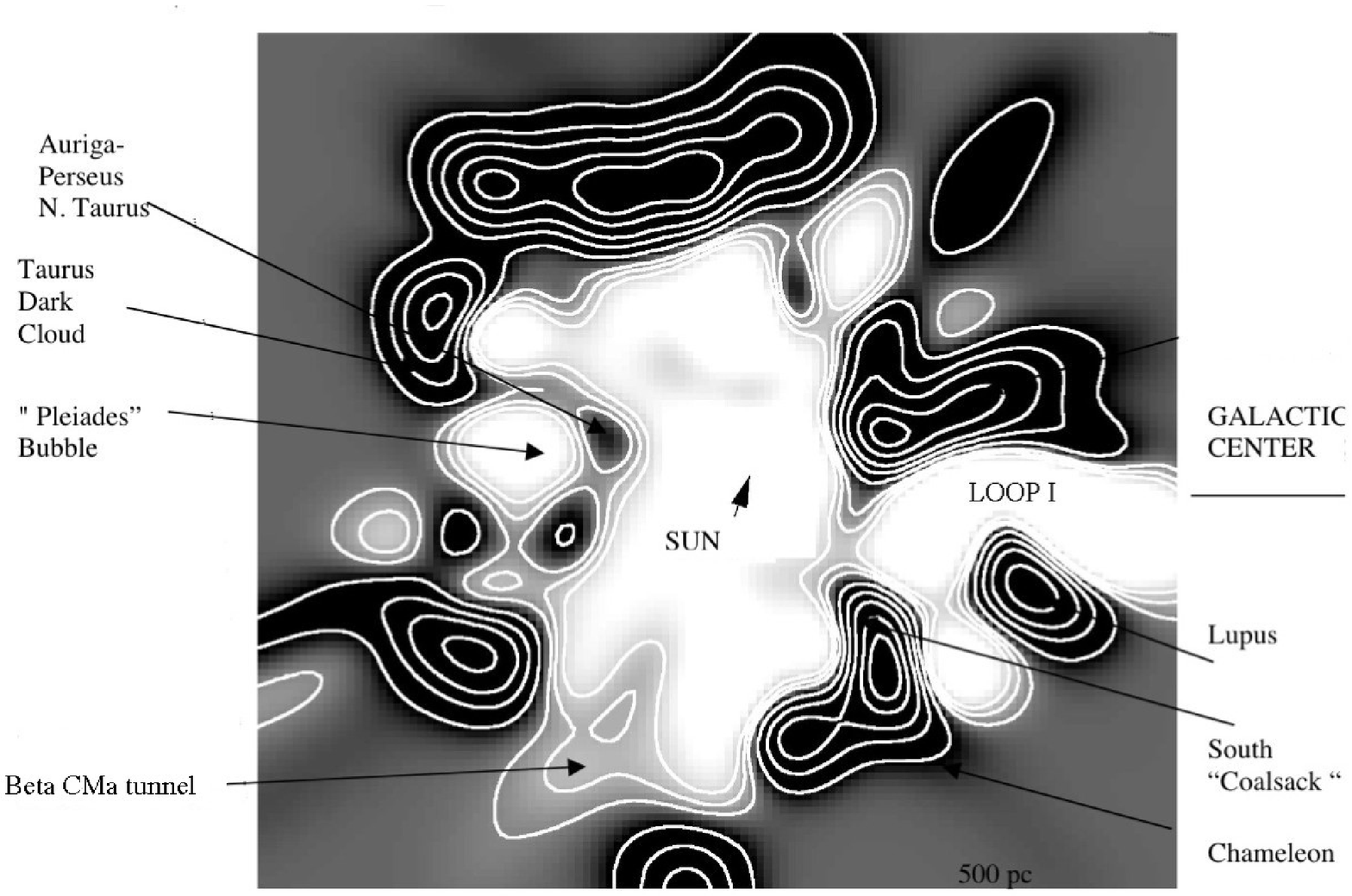}}
\caption{The distribution of cold and neutral gas within $\pm$250pc of the Sun in the galactic plane as revealed by interstellar NaI absorption measurements \citep{lall03}. Dark contour regions represent neutral gas with log N(HI) $>$ 19.3 cm$^{-2}$; white regions are of low gas density typically of log N(HI) $<$ 18.3 cm$^{-2}$. This is a view looking down onto the galaxy with the Sun at the center of the plot. Note the narrow extensions of the low density Local Cavity into surrounding interstellar cavities such as the Beta CMa tunnel, the Pleiades and Loop I bubbles.}
\end{figure}
\newpage

\begin{figure}
\center
{\includegraphics[height=8cm]{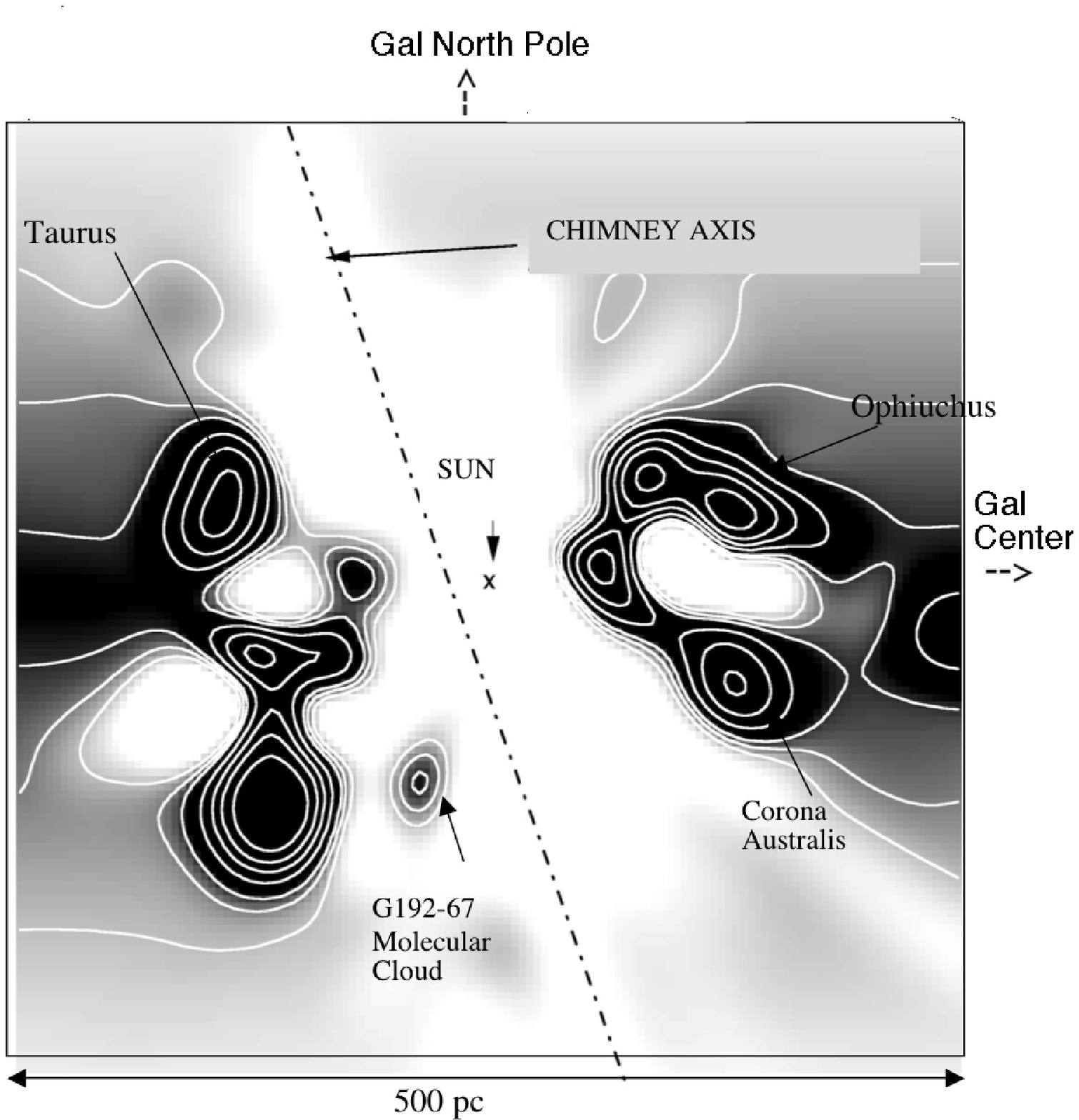}}
\caption{The distribution of cold and neutral gas within $\pm$250pc of the Sun in the plane perpendicular to the galactic plane as revealed by interstellar NaI absorption measurements \citep{lall03}. Dark contour regions represent neutral gas with log N(HI) $>$ 19.3 cm$^{-2}$; white regions are of low gas density typically of log N(HI) $<$ 18.3 cm$^{-2}$. This is a view looking side-on to the galaxy with the Sun at the center of the plot. Note the extension of the Local Cavity into the overlying  halo via the low density Local Chimney feature, which is tilted at an angle that points towards the regions of highest soft X-ray background emission.}
\end{figure}

\newpage

\begin{figure}
\center
{\includegraphics[height=9cm]{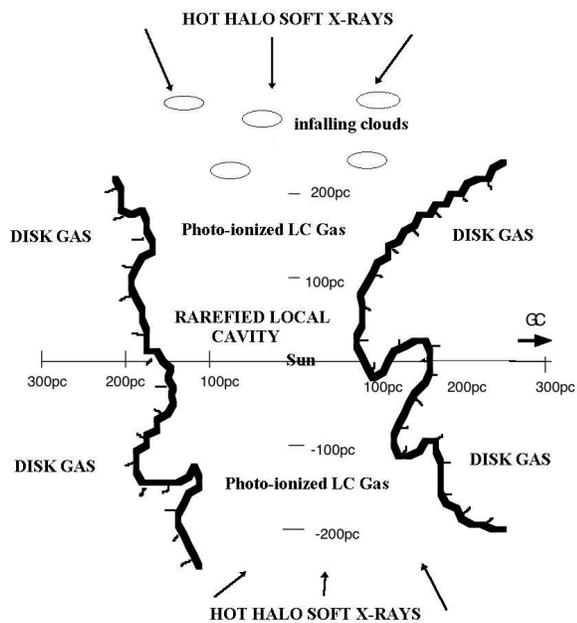}}
\caption{Simple 3-D schematic of the `Hot Top' Model. The figure shows the
spatial distribution of low neutral density photo-ionized gas contained within the Local Cavity surrounded
by its cold and neutral gas boundary wall and its openings into the halo through the Local Chimney as viewed in
the meridian plane (Lallement et al. 2003).
The entire region is subject to incident soft X-ray emission, arising in a million degree hot halo, that impinges on infalling gas clouds at high galactic latitudes.}
\end{figure}

\newpage

\begin{figure}
\center
{\includegraphics[height=5.5cm]{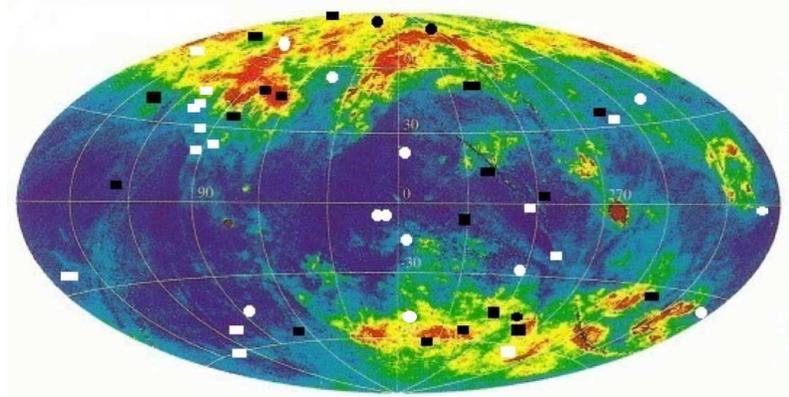}}
\caption{Spatial distribution of local \oxysix\ absorption towards local hot white dwarfs (Barstow
et al 2009, Savage and Lehner 2006) and B-type stars within 120pc (Bowen et al. 2008, Welsh $\&$ Lallement 2008) superposed
on a map of the distribution of the soft X-ray background
 R12 emission (Snowden et al. 1998). 
Sight-lines with
distances $<$ 80pc are circles, sight-lines with distances $>$ 80pc are squares. Filled black
circles or squares are detections of \oxysix, whereas non-detections are filled white circles
or squares. Note  that there are no detections of interstellar \oxysix\ along sight-lines $<$ 80pc that
are located near to the galactic plane.   Also note that 
the \oxysix\ detections are well correlated spatially with the regions of highest soft X-ray background emission intensity at highly positive or negative galactic latitudes. The figure has been adapted from Barstow et al. (2009).}

\end{figure}

\end{document}